\documentstyle[12pt,epsf]{article}
\begin{document}
\newcommand{\nd}[1]{/\hspace{-0.5em} #1}
\begin{titlepage}
\begin{flushright}
July 1992  \\
\end{flushright}

\begin{centering}
\vspace{.6in}
{\Large {\bf Vortex condensation in a model of random $\phi^{4}$-graphs}}

\vspace{.5in}
 N. Dorey and P. S. Kurzepa

\vspace{.1in}
Theoretical
Division, T-8 MS B285, Los Alamos National Laboratory, \\
Los Alamos, NM
87545, USA. \\
\vspace{.8in}
{\bf Abstract} \\
\vspace{.05in}
\end{centering}
{\small We consider a soluble model of large $\phi^{4}$-graphs
randomly embedded in one compactified dimension; namely the large-order
behaviour of finite-temperature
perturbation theory for the partition function of the anharmonic oscillator.
We solve the model using semi-classical methods and demonstrate the
existence of a critical
temperature at which the system undergoes a second-order phase transition
from $D=1$ to $D=0$ behaviour.
Non-trivial windings of the closed loops in a graph around the compactified
time direction are interpreted as vortices.
The critical point
has a natural interpretation as the temperature at which these vortices
condense and disorder the system. We show that the vortex density
increases rapidly in the critical region indicating the breakdown of
the dilute vortex gas approximation at this point.
We discuss the relation of this phenomenon to the
Berezinskii-Kosterlitz-Thouless transition in the $D=1$ matrix model
formulated on a circle.}
\end{titlepage}
\section{Introduction}
\paragraph{}
Models of random graphs have applications to theories of
polymers and random surfaces. An example of particular interest is the
$D=1$ Hermitian matrix model \cite{everyone,GK,kaz,KOG}
formulated on a circle of radius $R$ which is equivalent to
matrix quantum mechanics at a finite temperature $k_{B}T=1/2\pi R$.
The perturbative expansion of the matrix path-integral
for the case of a quartic potential
generates a sum over $\phi^{4}$ Feynman graphs,
each one dual to the quadrangulation of a random surface;
these surfaces being randomly embedded in $S^{1}$.
In the double-scaling limit, the resulting continuum theory can be
interpreted as an XY model coupled to two-dimensional quantum gravity or
as non-critical string theory in one compactified embedding dimension
(an extra dimension corresponding to the Liouville mode also arises).
An important
difference between a model of graphs embedded on the circle and one
formulated on the real line, is that the former admits vortex configurations
where a closed loop in the graph wraps around the compactified time
direction. Gross and Klebanov \cite{GK} have argued that these vortices are
suppressed at large radius but, at critical value of the
radius $R_{c}$,
the system undergoes a Berezinskii-Kosterlitz-Thouless (BKT) \cite{KT}
transition to a high-temperature phase in which
the vortices condense and disorder the embedding coordinate. They find that
the
embedding coordinate is completely randomized for $R<R_{c}$ and effectively
decouples leading to zero-dimensional behaviour in the high-temperature
phase.
\paragraph{}
Unfortunately the $D=1$ matrix model on a circle cannot be solved in the
double scaling limit by the usual methods which are effective in the case of
a non-compact target space. The reason is that the integral over the
angular degrees of freedom
of the matrix field, which can be done exactly on the real line to yield a
quantum mechanical problem involving the eigenvalues only, is no longer
tractable on the circle. Indeed this difficulty seems to be precisely
due to the
presence of vortices in the latter model: at low temperature
the model can be solved approximately
by considering only the $U(N)$ singlet sector, however this appears to be
equivalent to neglecting the vortex contribution
\footnote{Boulatov and Kazakov \cite{BK} have considered the adjoint sector
of the theory and have shown
that this corresponds to the contribution of a single
vortex-antivortex pair.}.
\paragraph{}
Similar vortex configurations occur in any model of
random graphs embedded on a circle. In general, these graphs need not
correspond directly to discretized random surfaces as they do for the matrix
model. In this letter we present such a model which exhibits
vortex condensation accompanied by an abrupt
transition from $D=1$ to $D=0$ behaviour and has the additional virtue of
being exactly soluble.  The model involves an ensemble of large
$\phi^{4}$-graphs randomly embedded
in a single compactified dimension and is
equivalent to the large-order behaviour of {\em finite-temperature}
perturbation theory for the
anharmonic oscillator. The graphs we consider carry no index structure, which
means that there is no unique prescription for attaching placquettes and
thereby defining a discretized surface. Thus, it will not be possible to
draw any definite
conclusions for one-dimensional string theory from our results.
Nevertheless, the model exhibits a non-trivial critical behaviour which
is similar to that expected at the BKT transition point of the $D=1$
matrix model and therefore merits study.
\paragraph{}
The large-order behaviour of perturbation theory for the quantum anharmonic
oscillator defined by the Hamiltonian,
\begin{equation}
H=(p^{2} + \phi^{2})/2 + g \phi^{4},
\label{ham}
\end{equation}
is determined by the instanton
solution \cite{LO,ZJ}, which is the classical motion
of a particle in the inverted potential
$V=-x^{2}/2 + x^{4}/4$ (see Figure 1).
At finite temperature
$T$, the relevant classical motion is
constrained to have period
$\beta=1/k_{B}T$. It is clear that there is a minimum period for which
there exists a non-trivial instanton
solution $x(t)$. The period is that of an infinitessimal
simple harmonic motion about the quadratic minimum of $V$ and is
determined by the curvature of the potential well at that point. Below this
critical period, the large-order behaviour of perturbation theory is governed
by the trivial solutions $x(t)=\pm 1$.
Although this phenomenon has no
special signifigance for the anharmonic oscillator itself
\footnote{The occurence of the minimum period was noted in this context
in \cite{ZJ}.},
we will argue that
it has a natural interpretation as a second-order phase transition in
the corresponding model of randomly embedded graphs. In particular,
we will show that the specific heat is
sharply peaked at the critical temperature and
exhibits a finite discontinuity at this point.
In addition we will show that
density of free vortices undergoes a crossover in the vicinity of the
critical point from a low temperature regime of tightly bound
vortex-antivortex pairs to
a high temperature regime populated by free (anti-)vortices. Thus we will
interpret the critical temperature as the point at which vortices condense
and disorder the system.
\paragraph{}
In Section 2, we define the partition function of the model as a sum over
Feynman graphs with a fixed large number of vertices
and identify the vortex configurations.
In Section 3 we solve the model using semi-classical methods
as described above and exhibit the critical behaviour. In the process, we
give some new analytic results for the large-order behaviour of
perturbation theory for the anharmonic oscillator at finite temperature.
Section 4 is devoted
to a discussion of these results.
\section{The model}
\paragraph{}
The partition function for the anharmonic oscillator defined by the
Hamiltonian (\ref{ham}),
at a finite temperature $T$, is given by
\begin{equation}
Z(\beta,g) =
{\rm Tr}\left[ e^{-\beta H}\right] =\int {\cal D}\phi \exp
\left[ -\int_{0}^{\beta} dt \frac{1}{2}
\dot{\phi}^{2} +\frac{1}{2}\phi^{2} + g \phi^{4} \right]
\label{zee}
\end{equation}
with $\beta=1/k_{B}T$.
Finite-temperature perturbation theory is generated by
expanding the
partition function as an asymptotic series in the coupling constant $g$,
\begin{equation}
Z(\beta,g)
\sim \sum_{k=0}^{\infty} Z_{k}(\beta) g^{k}
\label{pt}
\end{equation}
The coefficients $Z_{k}$ are given by the sum of all connected $\phi^{4}$
vacuum graphs $G$ with $k$ vertices and the Feynman rules
instruct us to assign a time coordinate $t_{i}\in [0,\beta]$
to each vertex and a finite-temperature propagator,
\begin{equation}
D_{ij}=
\sum_{m=-\infty}^{\infty} \exp\left[-|t_{i}-t_{j} + m\beta|\right]
\label{prop}
\end{equation}
to each link $\langle ij \rangle$ and then integrate over each $t_{i}$. The
propagator $D_{ij}$ has a geometrical interpretation as a sum over the
topologically inequivalent ways of embedding the link $\langle ij \rangle$
in the $S^{1}$
obtained by identifying the endpoints of the time interval
$[0,\beta]$ (see Figure 2). Each embedding of a link
$\langle ij \rangle$, is weighted by a factor $\exp ( -l \beta)$
in the summation
(\ref{prop}),
where $l$ is the length (in units of $\beta$)
of the image of $\langle ij \rangle$
in the target space.
\paragraph{}
Applying the Feynman rules and
summing over all graphs with $k$ vertices we have,
\begin{equation}
Z_{k}(\beta)=(-1)^{k}\sum_{G} S(G) \int_{0}^{\beta} \ldots \int_{0}^{\beta}
\prod_{i=1}^{k} dt_{i} \prod_{\langle ij \rangle} \sum_{m_{ij}=-\infty}^{
\infty}
\exp\left[ -|t_{i} - t_{j} + m_{ij}\beta|\right]
\label{part}
\end{equation}
For each graph $G$, a choice of the time coordinate
$\{t_{i}\}$ for each vertex and of a winding number $m_{ij}$ for each link
specifies an embedding of $G$ in $S^{1}$.
The coefficient $Z_{k}$ is a sum/integral over all such embeddings
of all $k$-vertex
vacuum graphs $G$, each one weighted by the symmetry factor $S(G)$
and by $\exp{(-L\beta)}$
where $L$ is the
total length of the image of $G$ in the target space.
In what follows, we will
consider $Z_{k}(\beta)$ itself, for fixed large $k$,
as defining the partition function of a theory of graphs
randomly embedded in $S^{1}$,
where the ``action'' of a ``configuration'' (ie a specific embedding)
is just its target-space length.
{}From this point of view,
$k\rightarrow\infty$ is the bulk limit of the model, and we can
define the usual thermodynamic quantities. In this limit,
the number of graphs grows as
$n_{k}\sim (16)^{k}(k-1)!$ and we define a normalized
partition function $\hat{Z}_{k}=Z_{k}/n_{k}$ which can be thought of as the
contribution of a typical graph.
The free energy per vertex is defined as,
\begin{equation}
f=-\lim_{k\rightarrow \infty} \frac{1}{k} \log |\hat{Z}_{k}(\beta)|
\label{free}
\end{equation}
and the corresponding specific heat is given by
$c_{v}=-\beta^{2}\partial^{2} f/\partial \beta^{2}$.
In Section 3 we will give
the exact solution of the model in the thermodynamic limit
by applying the standard semi-classical analysis
of large orders in perturbation theory to this finite temperature case.
In particular,
we will show that $f$ and $c_{v}$ remain finite as $k\rightarrow
\infty$ and that the latter
exhibits a discontinuity as a function of temperature indicating
a second-order phase transition.
It is convenient to express the partition function, $Z_{k}(\beta)$,
as a Laplace transform,
\begin{equation}
Z_{k}(\beta)=\int_{0}^{\infty} dL \tilde{Z}_{k}(L) \exp\left[ -L\beta \right]
\label{Laplace}
\end{equation}
The inverse Laplace transform, $\tilde{Z}_{k}(L)$, is the
corresponding microcanonical partition function for configurations of fixed
target-space length $L$.
\paragraph{}
For each configuration, it is possible to assign a vortex number $m_{\ell}$
to every closed loop $\ell$ in $G$. Specifically,
each term obtained by expanding the
product of propagators,
\begin{equation}
\prod_{\langle ij \rangle \in \ell} \sum_{m_{ij}= -\infty}^
{\infty} \exp\left[-|t_{i} - t_{j} + m_{ij} \beta|\right]
\end{equation}
is characterized by a vortex number $m_{\ell}=\sum_{\langle ij \rangle
\in \ell}
m_{ij}$ which labels
the homotopy class of the corresponding embedding of $\ell$ in
$S^{1}$. Negative values of $m_{\ell}$
correspond to anti-vortices and,
by analogy with the two-dimensional XY-model, we interpret loops for which
$m_{\ell}=0$ as tightly-bound vortex-antivortex pairs.
The total vortex number of a configuration is given by summing over all closed
loops in $G$, $M=\sum_{\ell} m_{\ell}$.
Because the sum over configurations (\ref{part})
is symmetric under $m_{ij}\rightarrow
-m_{ij}$, vortices and anti-vortices occur with equal statistical
weight and therefore $\langle
 M \rangle=0$.
\paragraph{}
For low temperatures, the summation for
each propagator in (\ref{part}) is
dominated by the $m_{ij}=0$ term. An isolated
(anti-)vortex necessitates at least
one of the $m_{ij}$ being non-zero and is thus suppressed by a factor of
$\exp{(-\beta)}$. Thus, at low temperature, we expect that
the links with non-zero $m_{ij}$ in a
typical configuration are
well seperated and therefore that
most vortices and antivortices are tightly bound in pairs.
The unbinding of a
vortex-antivortex pair is accompanied by
an increase in the target-space length
of the configuration by two units.
Thus, at least in this dilute low-temperature ground state,
the expected number of free (anti-)vortices
increases linearly with the expectation value of the quantity $L$.
In addition, when the corresponding length per vertex
$\rho=\langle L  \rangle/k$ becomes $O(1)$, it follows that
the free vortices have
become dense and the dilute gas approximation is no longer valid. In this
sense we will refer to $\rho$ as a ``vortex density''.
For any given configuration,
$L$ is also the corresponding action and so
the vortex density in the thermodynamic limit is given
by the internal energy per vertex,
$\rho=\partial f/\partial \beta$ (this formula can be obtained by
differentiating (\ref{Laplace})). In the next section,
we calculate $\rho$ explicitly and show that the system leaves the dilute
regime in the vicinity of the critical temperature. This suggests that
the phase transition
corresponds to a disordering of the system by the free vortices.
\section{Semi-classical solution for $k\rightarrow\infty$}
\paragraph{}
In this section we follow the standard approach to the large order behaviour
of perturbation theory for the anharmonic oscillator \cite{LO}
(for a pedagogical exposition see \cite{ZJ}).
Although this problem is frequently formulated
at finite temperature as a way of regulating the
infra-red divergences which occur at zero temperature, the explicit
evaluation of the temperature dependence of large-order behaviour
given below appears to be new.
The coefficients $Z_{k}(\beta)$ in (\ref{pt})
can be expressed in terms of the imaginary part of the
original partition function (\ref{zee}) for the anharmonic oscillator as,
\begin{equation}
Z_{k}(\beta)=\frac{1}{\pi}\int_{-\infty}^{0} dg \frac{ {\rm Im} Z(g)}{g^{k+1}}
\label{disp}
\end{equation}
reflecting the existence of a cut in the complex $g$-plane
along the negative real axis.
For $k \rightarrow \infty$ the integral is dominated by the
$g\rightarrow 0$
region of the integrand. In this region, the functional integral (\ref{zee})
for $Z(g)$ can be
evaluated accurately in the saddle-point approximation. The saddle-point
solutions, $x(t)=2\sqrt{-g}\phi(t)$,
are given by the appropriate periodic
classical trajectories
of a particle moving in the potential,
$V=-x^{2}/2 + x^{4}/4$
with turning points $x_{-}\leq x_{+}$ and total energy
$E=V(x_{\pm})$ (see Figure 1).
At finite
temperature the relevant trajectories have period
$\beta$. As discussed in the Introduction, there is a minimum period
$\beta_{c}$ for which a non-trivial solution exists. Expanding the potential
in the vicinity of the positive minimum $x_{c}=1$ we find,
\begin{equation}
V= -\frac{1}{4} + \frac{1}{2}\omega^{2}(x -
x_{c})^{2}
 + O((x-x_{c})^{3})
\label{e1}
\end{equation}
with $\omega=\sqrt{2}$. Thus as the total
energy of the particle approaches a critical value $E_{c}=-1/4$, its motion
becomes an infinitessimal harmonic oscillation about the well-bottom with
period $\beta_{c}=2\pi/\omega=\sqrt{2}\pi$.
Although $n$-instanton corrections corresponding to paths with periods
$\beta/n$ can occur at low temperature, these solutions do not occur for
$\beta<2\beta_{c}$.
\paragraph{}
The relevant instanton solutions, $x(t)$
satisfy the classical equation of
motion,
\begin{equation}
\ddot{x} -x +x^{3}=0
\label{ceq}
\end{equation}
with boundary conditions
$x(0)=x(\beta)$. For $\beta>\beta_{c}$, there are two
one-parameter families of such solutions generated by time-translations and
reflections, $x\rightarrow -x$. For definiteness, we will consider the
particular solution such that
$x(0)>0$ is the turning point of the motion $x_{-}$.
This solution is given as a
Jacobian elliptic function \cite{WW},
\begin{equation}
x(t)= x_{-} {\rm dn}\left[\frac{x_{-}t}
{\sqrt{2}}, k \right]
\label{csol}
\end{equation}
with imaginary modulus, $k=\sqrt{1-x^{2}_{+}/x^{2}_{-}}$.
\paragraph{}
The resulting saddle-point expression for ${\rm Im} Z(g)$ depends on $\beta$
as \cite{ZJ},
\begin{equation}
{\rm Im} Z(g) \sim -\frac{\beta}{4} \left(\frac{1}{2\pi g}
\frac{\partial E}{\partial \beta}
\right)^{\frac{1}{2}} \exp \left( - \frac{I(\beta)}{g} \right)
\label{form}
\end{equation}
where $I(\beta)$ is the classical action of the trajectory, which is
conveniently determined by the relation
\begin{equation}
\frac{\partial I(\beta)}{\partial \beta} =-\frac{E(\beta)}{4}
\label{ii}
\end{equation}
together with the
condition $I(\beta_{c})=\beta_{c}/16$ for the action of the limiting
infinitessimal
harmonic motion. The pre-exponential
factor in (\ref{form})
is given by the determinant of gaussian fluctuations about the
saddle-point and is proportional to $1/\sqrt{g}$ reflecting the
contribution of the zero-mode corresponding to time-translation of
the instanton solution. Adopting the convention of \cite{ZJ},
all functional
determinants considered (see also
(\ref{hhh}) below) are normalized with respect to the zero coupling result.
\paragraph{}
The formulae (\ref{ii},\ref{form})
demand that we find the dependence of the turning
point energy $E=V(x_{\pm})$ on  the period $\beta$. In particular
we are interested in the behaviour of $E(\beta)$ and its derivatives in the
critical region $(\beta- \beta_{c})<<1$. The inverse function, $\beta(E)$,
can be expressed as a first integral of the classical equation of motion
(\ref{ceq}),
\begin{equation}
\beta(E)=2\int_{x_{-}}^{x_{+}} \frac{dx}{\sqrt{2(E-V(x))}}=
\int_{E_{c}}^{E}dV \frac{dW}{dV} \frac{1}{\sqrt{2(E-V)}}
\label{inv}
\end{equation}
where $W(E)=x_{+}-x_{-}$ is the width of the well at energy $E<0$. The above
relation (Abel's integral equation) can be solved for $W$
by noting that the second integral in (\ref{inv})
is proportional
to the Riemann-Liouville integral\footnote{This integral can also
be thought of as arising from the standard Duhamel integral \cite{CH}
by explicit differentiation.} \cite{OS} which defines the fractional
derivative $D^{-\frac{1}{2}}$.
In particular, using the composition rule for fractional derivatives \cite{OS}
$D^{\delta}D^{\gamma}=D^{\delta+\gamma}$, we have
$\beta (E)\propto (D^{-\frac{1}{2}}D W)(E)=(D^{\frac{1}{2}}W)(E)$,
which implies that $W(E)\propto (D^{-\frac{1}{2}}\beta)(E)$ or,
reintroducing numerical factors,
\begin{equation}
W(E)=\frac{1}{\pi}\int_{E_{c}}^{E}dE' \frac{\beta(E')}{\sqrt{2(E-E')}}
\label{width}
\end{equation}
However, in the present case of a quartic potential, $W(E)$ is given
explicitly by,
\begin{eqnarray}
W(E)&=&\sqrt{1+\sqrt{E-E_{c}}}-\sqrt{1-\sqrt{E-E_{c}}} \nonumber \\
& = & 2\sum_{n=0}^{\infty} \left( {\frac{1}{2}} \atop {2n + 1} \right)
(E-E_{c})^{n+\frac{1}{2}}
\label{w2}
\end{eqnarray}
Comparing (\ref{width}) and (\ref{w2}) we find that $\beta(E)$ is
analytic for all $E<0$ and has a Taylor series,
\begin{equation}
\beta(E)=\beta_{c} +\beta_{c}\sum_{n=1}^{\infty} a_{n} (E-E_{c})^{n}
\label{exx}
\end{equation}
with coefficients,
\begin{equation}
a_{n}=\left({\frac{1}{2}} \atop 2n +1 \right) \bigg{/}
\sum_{k=0}^{n}\left(n \atop k \right) \frac{(-1)^{k}}{2k + 1}
\label{coef}
\end{equation}
\paragraph{}
Standard theorems on implicit functions imply that $E$ is an
infinitely differentiable function of $\beta$ in some neighbourhood of
$\beta_{c}$. The Taylor series for
$E(\beta)$ in
powers of $(\beta-\beta_{c})$ can be derived by inverting the
series (\ref{exx}). Although the general term in this series cannot be
expressed in closed form, the
coefficients can be generated systematically
using the algorithm described in the appendix. For the present purpose,
we will require only the leading terms,
\begin{equation}
E(\beta)= -\frac{1}{4} + \frac{4}{3\beta_{c}}(\beta - \beta_{c}) +
O((\beta - \beta_{c})^{2})
\label{e2}
\end{equation}
For $\beta\rightarrow \infty$, $E(\beta)$ has a double expansion in powers of
$\beta$ and $e^{-\beta}$. A straightforward calculation yields,
\begin{equation}
E(\beta)=-16 e^{-\beta} - 64(3\beta - 10)e^{-2\beta} +
O(\beta^{2}e^{-3\beta})
\label{betainf}
\end{equation}
The first term in this series yields the standard zero-temperature result
\cite{ZJ} when substituted for $E$ in (\ref{form}).
We computed $E(\beta)$ numerically for the whole range $\beta_{c}<\beta
<\infty$ by solving equation (\ref{inv}), the
resulting graph is shown in Figure 3. We also computed $I(\beta)$ by
integrating our numerical solution for $E(\beta)$.
\paragraph{}
For $\beta<\beta_{c}$, the only solutions of the equation of motion are the
trivial ones $x(t)=\pm 1$. The corresponding contribution to the
partition function is,
\begin{equation}
{\rm Im}Z(g)\sim
2\left(\frac{\sqrt{2}\sinh{\beta}}{\sinh{\sqrt{2}\beta}}\right)
^{\frac{1}{2}} \exp\left(- \frac{\beta}{16g}\right)
\label{hhh}
\end{equation}
where the determinant prefactor is independent of $g$ as the trivial
saddle-point has no zero-mode. Applying equation (\ref{disp}), we find that
$\beta_{c}$ separates two ``phases'' characterized by
different asymptotic
behaviour of $Z_{k}$. For $\beta>\beta_{c}$ we have,
\begin{eqnarray}
Z_{k}& \sim & (-1)^{k}\frac{\beta}{2}\left(\frac{1}{2\pi}
\frac{\partial E}{\partial \beta}
\right)^{\frac{1}{2}}\left(I(\beta)
\right)^{-(k +\frac{1}{2})}\Gamma \left(k + \frac{1}{2}\right)
\label{lorder1}
\end{eqnarray}
whereas for $\beta<\beta_{c}$,
\begin{eqnarray}
Z_{k} & \sim  & (-1)^{k}
2\left(\frac{\sqrt{2}\sinh{\beta}}{\sinh{\sqrt{2}\beta}}
\right)^{\frac{1}{2}}\left(\frac{\beta}{16}\right)^{-k}\Gamma (k)
\label{lorder}
\end{eqnarray}
In the low-temperature phase, $Z_{k}$ exhibits the same
$I^{-k}k^{1/2}(k-1)!$ growth, as the
zero-temperature result. In contrast, the large-order behaviour of
perturbation theory in the high-temperature phase is
identical (up to a $k$-independent prefactor) to that of the perturbative
expansion of the ordinary integral,
\begin{equation}
Z^{*}=\int_{0}^{\infty} dy \exp
\left[-\frac{\beta}{4}\left(\frac{y^2}{2} +g y^4\right)
\right]
\label{zeezero}
\end{equation}
In this sense, the large-order behaviour of perturbation theory
exhibits a transition from $D=1$ to $D=0$ behaviour at $\beta=\beta_{c}$.
\paragraph{}
Equations (\ref{lorder1}) and (\ref{lorder}) constitute an exact solution for
the partition function of the
model of random graphs considered in Section 2 in the thermodynamic limit.
Although the partition function itself grows wildly with $k$ reflecting the
$k!$ growth in the number of Feynman diagrams,
the free energy per vertex of a typical graph (\ref{free})
remains finite as does the corresponding specific heat. For
$\beta>\beta_{c}$ we find,
\begin{eqnarray}
c_{v} & = & \frac{\beta^{2}}{4}\left[\frac{\partial E}{\partial \beta}
\frac{1}{I(\beta)} + \frac{E^{2}(\beta)}{4 I^{2}(\beta)} \right]
\label{therm1}
\end{eqnarray}
while for $\beta<\beta_{c}$, $c_{v}=1$. Using equations (\ref{ii},
\ref{e2}, \ref{therm1}) we find that $c_{v}\rightarrow 19/3$ as
$\beta\downarrow\beta_{c}$.
The resulting discontinuity in the specific
heat at $\beta=\beta_{c}$ implies that the system undergoes a second-order
phase transition at this point. The result of a
numerical evaluation of $c_{v}$ is shown in Figure 4. As expected for
a second-order phase transition, the specific heat has a narrow peak in the
critical region.
\paragraph{}
In Section 2 we argued that the density of free vortices is given by
the expectation value of the target-space length per vertex $L/k$ which is
exactly the internal energy $\partial f/\partial \beta$.
Hence, for $\beta>\beta_{c}$, we have $\rho=-E(\beta)/4I(\beta)$
while for $\beta<\beta_{c}$, $\rho=1/\beta$. The corresponding graph of
the vortex density against temperature is shown in Figure 5.
At low temperature, the vortex density is exponentially suppressed,
\begin{equation}
\rho\sim \exp \left(-\frac{1}{k_{B}T}\right)
\label{dilute}
\end{equation}
This implies that the low-temperature ground state of the system can
be thought of as a dilute gas of tightly-bound vortex-antivortex pairs.
In the high-temperature phase the vortex density is $O(1)$ and
increases linearly with
temperature indicating a breakdown of the dilute gas approximation.
The numerical results shown in Figure 5 reveal
a rapid crossover between
these two regimes in
a narrow region near the critical temperature $k_{B}T_{c}=1/\sqrt{2}\pi$.
In particular, the vortex density increases by a factor of about $6.35$
in the interval $0.75 T_{c}<T<T_{c}$. This supports an
interpretation of the critical temperature as the point at
which vortex-antivortex pairs unbind and disorder the system. There is a
qualitative similarity between our numerical results for $\rho$
and an
evaluation of the vortex density in a Monte-Carlo simulation of the
two-dimensional XY model presented in \cite{rajan}.
\section{Conclusions}
\paragraph{}
In this paper we have presented a novel interpretation for the large-order
behaviour of finite-temperature perturbation theory. In particular
we have shown that a fixed large order in the perturbation series for the
partition function of an anharmonic oscillator defines a non-trivial
statistical mechanical model with a sensible thermodynamic limit.
The model exhibits two phases separated by a
second-order phase transition and it is important to establish the
universality class to which this critical behaviour
belongs. Although this transition has the same interpretation
as the BKT transition in the two-dimensional XY model, the latter is a
transition of infinite order having
no discontinuities in the derivatives of the free energy.
\paragraph{}
The behaviour of $Z_{k}$ for $\beta<\beta_{c}$
is identical to the large-order behaviour of the perturbative expansion of an
ordinary integral. The perturbation series for the integral $Z^{*}$ is just
a sum over all $\phi^{4}$ graphs with no embedding dimension. This suggests
that,
in the high-temperature phase of the model, the embedding coordinate is
completely randomized by the gas of free vortices and
decouples leaving an effectively zero-dimensional theory. This is exactly
the behaviour expected at the BKT transition point of the $D=1$ matrix model
\cite{GK}. However, due to the unconventional procedure of summing over
all Feynman graphs with equal weight irrespective of genus,
there is no direct correspondence between our model and
the double-scaling limit of the $D=1$ matrix model. In addition,
we note that the estimates for the
critical temperature of the matrix model given in \cite{GK,BK}
differ from the critical temperature found here by a factor of two.
On the other hand,
the occurence of vortices is a purely
local phenomenon on any
graph or discretized surface and it is expected that the
BKT transition should occur independently at each order in the genus
expansion. Hence one might expect the same critical behaviour
to appear even in the
unweighted sum over different genera considered here. Thus it is
possible that our model exhibits the
same universal behaviour as the XY model coupled to two-dimensional
quantum gravity.
\paragraph{}
The model presented in this paper has several unusual features. As in the
matrix models, the sum over configurations is generated by the
perturbative expansion of a functional integral and can be evaluted using
semi-classical methods. However, in our case, the thermodynamic limit of
large graphs is taken explicitly rather than by tuning the coupling constant
to a critical value (indeed the latter procedure is not available here
as the perturbation series has zero radius of convergence). In principle
we can obtain a continuum theory only at the critical temperature
$\beta=\beta_{c}$. Finally we note that the model considered here can be
generalised to any theory where the large-order behaviour of perturbation
theory may be evaluated by saddle-point methods.
\paragraph{}
The authors acknowledge useful discussions with Jan Ambjorn,
Paul Ginsparg, Tomek Jaroszewicz, and Ian Kogan. PSK thanks the
Aspen Center for Physics,
where part of this work was completed, for its hospitality.

\section*{Appendix}
\paragraph{}
In this appendix we give a general
formula for the inversion of a power series of
the form,
\begin{equation}
w=f(z)=\sum_{n=0}^{\infty} a_{n}z^{n}
\label{psa}
\end{equation}
which provides an algorithm for generating the coefficients of the series
(\ref{e2}) for $E(\beta)$. We consider the inversion of (\ref{psa}) in an
open neighbourhood of $z=0$; accordingly we demand that $f'(0)\neq 0$ and set
$f(0)=w_{0}$. The inverse series for $z(w)$ can be written as,
\begin{equation}
z= \frac{1}{a_{1}}(w-w_{0})+
\sum_{m=2}^{\infty}\frac{(w-w_{0})^{m}}{m!}\frac{(-1)^{m+1}(m-2)}
{a_{1}^{2m-1}}
\sum_{l_{1}=0}^{\infty}\cdots \sum_{l_{m}=0}^{\infty}\delta_
{(\sum l_{i}-m+1, 0)} \quad{}C_{l_{1}}\ldots C_{l_{m}}
\label{bast}
\end{equation}
where
\begin{equation}
C_{l}=\left|\begin{array}{ccccc} a_{2}  & a_{1} & 0 &\ldots & 0 \\
a_{3} & a_{2} & a_{1} & \ldots & 0  \\
\vdots & \vdots & \vdots & \ddots & \vdots \\
a_{l} & a_{l-1} & a_{l-2} & \ldots & a_{1} \\
a_{l+1} & a_{l} & a_{l-1} & \ldots & a_{2} \end{array} \right|
\end{equation}
More explicitly, if
\begin{equation}
w=w_{0}+az+bz^{2}+cz^{3}+dz^{4}+\ldots
\end{equation}
(\ref{bast}) gives,
\begin{equation}
z=\frac{1}{a}(w-w_{0})-\frac{b}{a^{3}}(w-w_{0})^{2}-
\frac{1}{a^{5}}(ac-2b^{2})(w-w_{0})^{3}-\frac{1}{a^{7}}(a^{2}d-5abc+5b^{3})
(w-w_{0})^{4}+\ldots
\end{equation}
\section*{Figure Captions}
\paragraph{}
Figure 1. The potential $V(x)$.
\paragraph{}
Figure 2. The geometrical interpretation of the finite-$T$ propagator
(\ref{prop}) as a sum over topologically inequivalent embeddings of the link
$\langle ij \rangle$ in $S^{1}$.
\paragraph{}
Figure 3. A graph of $E$ as a function of $\beta/\beta_{c}$.
\paragraph{}
Figure 4. A graph of $c_{v}$ as a function of $k_{B}T=1/\beta$.
\paragraph{}
Figure 5. A graph of $\rho$ as a function of $k_{B}T$.

\begin{thebibliography}{99}
\bibitem{everyone} V. A. Kazakov and A. Migdal, {\em Nucl. Phys.} {\bf B311}
(1989) 171.
\par D. J. Gross and  N. Miljkovic, {\em Phys. Lett.}
{\bf B238} (1990) 217.
\par E. Brezin, V. A. Kazakov and Al. B. Zamolodchikov, {\em Nucl. Phys.}
{\bf B338} (1990) 673.
\par P. Ginsparg and J. Zinn-Justin, {\em Phys. Lett.} {\bf B240} (1990) 333.
\bibitem{GK} D. J. Gross and I. R. Klebanov, {\em Nucl. Phys.} {\bf B344}
(1990) 475 and {\em Nucl. Phys.} {\bf B354} (1991) 459.
\bibitem{kaz} V. A. Kazakov, LPTENS preprint 90/30, to appear in ``Random
Surfaces and Quantum Gravity'' ed. by O. Alvarez et. al.
\par Z. Yang, {\em Phys. Lett.} {\bf B243} (1990) 365.
\bibitem{KOG} I. I. Kogan, {\em Phys. Lett.} {\bf B281} (1991) 233.
\bibitem{KT} V. L. Berezinskii, {\em JETP} {\bf 34} (1972) 610.
\par M. Kosterlitz and D. Thouless, {\em J. Phys.} {\bf C6} (1973) 1181.
\bibitem{BK} D. V. Boulatov and V. A. Kazakov, LPTENS preprint 91/24.
\bibitem{LO} C. M. Bender and T. T. Wu, {\em Phys. Rev. } {\bf D7} (1973) 1620.
\par E. Brezin, J. C. Le Guillou and J. Zinn-Justin, {\em Phys. Rev.}
{\bf D15} (1977) 1544 and {\em Phys. Rev.} {\bf D15} (1977) 1558.
\bibitem{ZJ} J. Zinn-Justin, ``Quantum Field Theory and Critical
Phenomena'' (Clarendon Press 1989) pp 836 and
{\em Phys. Rep.} {\bf 70} (1981) 109.
\bibitem{WW} E. T. Whittaker and G. N. Watson ``A Course of Modern Analysis''
(Cambridge University Press 1943).
\bibitem{OS} K. B. Oldham and J. Spanier, ``The Fractional Calculus''
(Academic Press 1975).
%
\bibitem{CH} R. Courant and D. Hilbert, ``Methods of Mathematical Physics'',
vol 2 (John Wiley 1989).
%
\bibitem{rajan} R. Gupta and C. F. Baillie, {\em Phys. Rev.} {\bf B45} (1992)
2883.



\end{thebibliography}
\end{document}